\documentclass[12pt]{iopart}
\usepackage{setstack}
\usepackage{graphicx}
\usepackage{amssymb}
\begin{document}
\title[Cosmography: Cosmology without the Einstein equations]
{Cosmography: \\ 
Cosmology without the Einstein equations}
\author{Matt Visser}
\address{School of Mathematics, Statistics, and Computer Science,\\ 
Victoria University of Wellington, PO Box 600, Wellington, New Zealand}
\ead{matt.visser@mcs.vuw.ac.nz}
\begin{abstract}
\def\d{{\mathrm{d}}}
How much of modern cosmology is really cosmography? How much of modern
cosmology is independent of the Einstein equations? (Independent of
the Friedmann equations?) These questions are becoming increasingly
germane --- as the models cosmologists use for the stress-energy
content of the universe become increasingly baroque, it behoves us to
step back a little and carefully disentangle cosmological kinematics
from cosmological dynamics. The use of basic symmetry principles (such
as the cosmological principle) permits us to do a considerable amount, without
ever having to address the vexatious issues of just how much ``dark
energy'', ``dark matter'', ``quintessence'', and/or ``phantom matter''
is needed in order to satisfy the Einstein equations. This is the
sub-sector of cosmology that Weinberg refers to as ``cosmography'',
and in this article I will explore the extent to which cosmography is
sufficient for analyzing the Hubble law and so describing many of the
features of the universe around us.

\vskip 0.50cm

\noindent
\emph{Based on a talk presented at ACRGR4, the 4th Australasian Conference
on General Relativity and Gravitation, Monash University, Melbourne,
January 2004.}

\noindent{To appear in the proceedings, in \bf{General Relativity and Gravitation}.}

\vskip 0.50cm
\noindent
  Dated: 14 June 2004; \LaTeX-ed \today
\end{abstract}
\maketitle
\newtheorem{theorem}{Theorem}
\newtheorem{corollary}{Corollary}
\newtheorem{lemma}{Lemma}
\def\d{{\mathrm{d}}}
\def\implies{\Rightarrow}

\def\eg{{\it e.g.}}
\def\etc{{\it etc}}
\def\sign{{\hbox{sign}}}
\def\eof{\Box}
\newenvironment{warning}{{\noindent\bf Warning: }}{\hfill $\eof$\break}
\section{Introduction}

In this article I discuss a ``phenomenological'' approach to
cosmography. Specifically I take a hard look at the question of just
how much of modern cosmology can be extracted from symmetry principles
and direct observation --- without ever invoking the Einstein equations
(Friedmann equation), and so without ever having to deal with
contentious issues regarding ``dark energy'', ``dark matter'',
``quintessence'', and/or ``phantom matter''~\cite{jerk-and-snap}.

Indeed, a surprising amount of modern cosmology is pure kinematics,
what Weinberg~\cite{Weinberg} refers to as cosmography, and is
completely independent of the underlying dynamics governing the
evolution of the universe.  For instance, it is well-known that basic
symmetry principles (and in particular the cosmological principle) are
sufficient to deduce the \emph{form} of the cosmological metric --- up
to possible topological ambiguities it must be a
Friedmann--Robertson--Walker [FRW] cosmology
\begin{equation}
\d s^2 = - c^2 \;\d t^2 
+ a(t)^2 \left[ {\d r^2\over 1- k \,r^2} 
+ r^2\left(\d\theta^2+\sin^2\theta\;\d\phi^2\right) \right].
\end{equation}
Whereas pure cosmography by itself will not predict anything about the
scale factor $a(t)$, in the cosmographic scenario we can to some
extent \emph{infer} the history of the scale factor $a(t)$ from the
observational data while steadfastly avoiding use of the Einstein
equations.  In view of the many controversies currently surrounding
the composition of the cosmological fluid, and the large number of
speculative models presently being considered, such an observationally
driven reconstruction is of interest in its own right.

\section{Hubble law}

Now in observational cosmology we do not have direct access to the
complete history of the scale factor $a(t)$ over the entire age of the
universe --- we do however have access [however imprecise] to the
current value of the scale factor and its derivatives, as encoded in
the Hubble parameter, deceleration parameter, \etc. This more limited
information can still be used to extract useful information about the
[relatively recent] history of our universe.

To set the notation as in reference~\cite{jerk-and-snap}, it is
standard terminology in mechanics that the first four time derivatives
of position are referred to as velocity, acceleration, jerk, and snap.
Jerk [the third time derivative] is also sometimes referred to as
jolt.  Less common alternative terminologies are pulse, impulse,
bounce, surge, shock, and super-acceleration. Snap [the fourth time
derivative] is also sometimes called jounce. The fifth and sixth time
derivatives are sometimes somewhat facetiously referred to as crackle
and pop.

So in a cosmological setting this makes it appropriate to define
Hubble, deceleration, jerk, and snap parameters as:
\begin{equation}
H(t) = + {1\over a} \; {\d a\over\d t};
\end{equation}
\begin{equation}
q(t) = - {1\over a} \; {\d^2 a\over \d t^2}  \;
\left[ {1\over a} \; {\d a \over  \d t}\right]^{-2};
\end{equation}
\begin{equation}
j(t) = + {1\over a} \; {\d^3 a \over \d t^3}  
\; \left[ {1\over a} \; {\d a \over  \d t}\right]^{-3};
\end{equation}
\begin{equation}
s(t) = + {1\over a} \; {\d^4 a \over \d t^4}  
\; \left[ {1\over a} \; {\d a \over  \d t}\right]^{-4}.
\end{equation}
The deceleration, jerk, and snap parameters defined in this way are
dimensionless, and we can write
\begin{eqnarray}
\fl
a(t)= a_0 \;
\Bigg\{ 1 + H_0 \; (t-t_0) - {1\over2} \; q_0 \; H_0^2 \;(t-t_0)^2 
+{1\over3!}\;  j_0\; H_0^3 \;(t-t_0)^3 
\nonumber
\\
\qquad
+{1\over4!}\;  s_0\; H_0^4 \;(t-t_0)^4
+ O([t-t_0]^5) \Bigg\}.
\end{eqnarray}

Now the physical distance travelled by a photon that is emitted at
time $t_*$ and absorbed at the current epoch $t_0$ is
\begin{equation}
D = c \int \d t = c\;(t_0 - t_*),
\end{equation}
where the time difference $\Delta t = t_0-t_*$ is called the
``lookback time''.  In terms of this physical distance travelled the
Hubble law is \emph{exact} but completely impractical:
\begin{equation}
1 + z = {a(t_0)\over a(t_*)} = {a(t_0)\over a(t_0 - \Delta t)} 
= {a(t_0)\over a(t_0 - D/c)},
\end{equation}
A more useful result is obtained by performing a Taylor series
expansion. Working to fourth order in $D$, or more precisely to fourth
order in the dimensionless parameter $D H_0/c$, yields
\begin{eqnarray}
\fl
{a(t_0)\over a(t_0 - D/c)} =
1 + {H_0 D\over c} +{2+q_0\over2} \; {H_0^2 D^2\over c^2} 
+ 
{6(1+q_0)+j_0\over6} \; {H_0^3 D^3\over c^3} 
\nonumber
\\
+{24-s_0+8j_0+36q_0+6q_0^2\over24} \;
 {H_0^4 D^4\over c^4} +
 O\left[\left(H_0 D\over c\right)^5\right].
\end{eqnarray}
So that
\begin{eqnarray}
z(D) &=&  {H_0 D\over c} +{2+q_0\over2} \; {H_0^2 D^2\over c^2} 
+ 
{6(1+q_0)+j_0\over6} \; {H_0^3 D^3\over c^3} 
\nonumber
\\
&&+{24-s_0+8j_0+36q_0+6q_0^2\over24} \;
 {H_0^4 D^4\over c^4} +
 O\left[\left(H_0 D\over c\right)^5\right].
\end{eqnarray}
Reversion of this power series, to convert $z(D)\to D(z)$, leads
to:
\begin{eqnarray}
\fl
D(z) = {c\; z\over H_0}
\Bigg\{ 1 
- 
\left[1+{q_0\over2}\right] {z} 
+
\left[ 1 + q_0 + {q_0^2\over2} - {j_0\over6}   \right] z^2 
\\
-
\left[ 1 +{3\over2}q_0(1+q_0)+{5\over8}q_0^3-{1\over2}j_0 
- {5\over12} q_0 j_0 -{s_0\over24} \right] z^3
+  O(z^4) \Bigg\}.
\nonumber
\label{E:physical-Hubble}
\end{eqnarray}
This simple calculation is enough to demonstrate that the jerk shows
up at third order in the Hubble law, and the snap at fourth order.
Generally, the $O(z^n)$ term in this version of the Hubble law will
depend linearly on the $n$-th time derivative of the scale factor, and
nonlinearly on lower-order time derivatives.  (Also note that one of
the virtues of this particular version of the Hubble law is that it is completely
independent of $k$, the sign of space curvature, and is completely
independent of $a_0$, the present-day value of the scale factor.)
Carrying out fifth-order or even sixth-order expansions in terms of
analogously defined crackle and pop parameters is straightforward with
the aid of a symbolic algebra system such as \emph{Maple}, but the
formulae grow so clumsy as to be not particularly useful.

Unfortunately physical distance $D$ (or equivalently the lookback time
$\Delta t$) is typically not the variable in terms of which the Hubble
law is observationally presented. That role is more typically played
by the ``luminosity distance'', $d_L$. For instance, Weinberg
defines~\cite{Weinberg}
\begin{equation}
\hbox{(energy flux)} = {L\over4\pi \; d_L^2}.
\end{equation}
Let the photon be emitted at $r$-coordinate $r=0$ at time $t_*$, and
absorbed at $r$-coordinate $r=r_0$ at time $t_0$. Then it is a purely
geometrical textbook result that
\begin{equation}
d_L = a(t_0)^2 \; {r_0\over a(t_*)} = {a_0\over a(t_0-D/c)} \; (a_0\,r_0).
\end{equation}
Thus to calculate $d_L(D)$ we need $r_0(D)$.  A brief and quite
standard computation yields
\begin{equation}
r_0(D) = \left\{
\begin{array}{cc}
\sin\left( \displaystyle\int_{t_0-D/c}^{t_0} {c\;\d t \over a(t)} \right) 
      & k=+1;\\
\\
\displaystyle\int_{t_0-D/c}^{t_0} {c\;\d t \over a(t)}        
                  & k=0;\\
\\
\sinh\left( \displaystyle\int_{t_0-D/c}^{t_0} {c\;\d t \over a(t)} \right) 
\;\; & k=-1;
\end{array}
\right.
\end{equation}
where we now must deal with the three possible signs for space
curvature, $k = -1/0/+1$, separately.  We now Taylor series expand
this result for ``short'' distances $D$. First note that
\begin{equation}
\fl
r_0(D) = 
\left[\int_{t_0-D/c}^{t_0} {c\;\d t \over a(t)}\right]
- {k\over 3!} 
\left[ \int_{t_0-D/c}^{t_0} {c\;\d t \over a(t)} \right]^3 
+ O\left( \left[ \int_{t_0-D/c}^{t_0} {c\;\d t \over a(t)} \right]^5 \right),
\end{equation}
and observe that the sign of the space curvature $k$ explicitly
shows up in the third-order term.  Now expand the integrals above to
third order. (We can easily check, \emph{a posteriori}, that this is
sufficient for the final result for $d_L(z)$ quoted below.)  Then
\begin{eqnarray}
\fl
\int_{t_0-D/c}^{t_0} {c\;\d t \over a(t)}
=
 {D\over a_0} \Bigg\{ 1 +{1\over2} {H_0 D\over c} 
+\left[ {2+q_0\over6} \right]
\left({H_0 D\over c}\right)^2
+\left[{6(1+q_0)+j_0\over24}\right]
\left({H_0 D\over c}\right)^3
\nonumber\\ 
+
O\left[\left(H_0 D\over c\right)^4\right]\Bigg\}.
\end{eqnarray}
So we see that the conversion from $D$, the physical distance travelled, to $r$
coordinate traversed is given by
\begin{eqnarray}
\fl
r_0(D) = {D\over a_0} 
\Bigg\{ 1 +{1\over2} {H_0 D\over c} + 
 {1\over6} \left[ 2+q_0 - {kc^2\over H_0^2 a_0^2} \right] \; 
\left(H_0 D\over c\right)^2 
\nonumber
\\
+
{1\over24} \left[6(1+q_0)+j_0  - 6 {kc^2\over H_0^2 a_0^2}\right] 
\left(H_0 D\over c\right)^3
+
O\left[\left(H_0 D\over c\right)^4\right]\Bigg\}.
\end{eqnarray}
Combining these formulae we find that the luminosity distance as a
function of $D$, the physical distance travelled, is:
\begin{eqnarray}
\fl
d_L(D) = {D}  \Bigg\{ 1  +{3\over2} \left({H_0 D\over c}\right) 
+
{1\over6}\left[11+4q_0-{kc^2\over H_0^2a_0^2}\right]
\left(H_0 D\over c\right)^2
\nonumber
\\
+
{1\over24}\left[50+40q_0+5j_0-10{k c^2\over H_0^2 a_0^2}    \right]
\left(H_0 D\over c\right)^3
+ O\left[\left(H_0 D\over c\right)^4\right]\Bigg\}.
\end{eqnarray}
Now using the series expansion for for $D(z)$ we finally derive, on
purely geometrical grounds, the luminosity-distance version of the
Hubble law:
\begin{eqnarray}
\fl
d_L(z) =  {c\; z\over H_0}
\Bigg\{ 1 + {1\over2}\left[1-q_0\right] {z} 
-{1\over6}\left[1-q_0-3q_0^2+j_0+ {kc^2\over H_0^2a_0^2}\right] z^2
\nonumber
\\
\lo{+}
{1\over24}\left[
2-2q_0-15q_0^2-15q_0^3+5j_0+10q_0j_0+s_0 + 
{2 k c^2(1+3q_0)\over H_0^2 a_0^2}
\right] z^3
\nonumber
\\
+ O(z^4) \Bigg\}.
\end{eqnarray}
The first two terms above are Weinberg's version of the Hubble law.
His equation (14.6.8). The third term is equivalent to that obtained
by Chiba and Nakamura~\cite{Chiba}, and by Visser~\cite{jerk-and-snap},
and depends on the jerk parameter $j_0$, the sign of space curvature
$k$, and the present day value of the scale factor $a_0$. It is only
at this third-order term in the Hubble law that we even begin to probe
the geometry of space, and even then the fact that we are sensitive to
the geometry of space depends on our choice of distance scale ---
recall that the physical distance Hubble law $D(z)$ as embodied in
equation (\ref{E:physical-Hubble}) is completely insensitive to the
geometry of space (not spacetime). The fourth-order term of either the
luminosity distance or the physical distance version of the Hubble law
is (as expected) linearly dependent on the snap. From the derivation
above it is now clear that the $O(z^n)$ term in this luminosity
distance version of the Hubble law will also depend linearly on the
$n$-th time derivative of the scale factor. It is also clear, if somewhat
awkward, how to extend the calculation to arbitrarily high order
in redshift.

If one instead chooses to work with angular diameter distance $d_A(z)$
or proper-motion distance $d_M(z)$ the relevant conversions are
straightforward~\cite{Weinberg}
\begin{equation}
d_A(z) = {d_L(z)\over(1+z)^2}; 
\qquad
d_M(z) = {d_L(z)\over1+z}. 
\end{equation}

It is important to realise that any of these versions of the Hubble
law, and indeed the entire discussion of this article, is completely
independent of the Einstein equations --- it assumes only that the
geometry of the universe is well approximated by a FRW cosmology, but
does not invoke any particular matter model.  Note further that in
comparison to the physical distance travelled $D(z)$ Hubble law, this
luminosity distance $d_L(z)$ Hubble law first differs in the
coefficient of the $O(z^2)$ term --- you will still get the same
Hubble parameter, but if you are not sure which definition of
``distance'' you are using you may mis-estimate the higher-order
coefficients (deceleration, jerk, and snap).

\section{Cosmological inflation}

From a theoretical perspective, $H_0 \, a_0/c \gg 1$ is a generic
prediction of inflationary cosmology --- thus assuming cosmological
inflation effectively permits is to write
\begin{eqnarray}
\fl
d_L(z) =  {c\; z\over H_0}
\Bigg\{ 1 + {1\over2}\left[1-q_0\right] {z} 
-{1\over6}\left[1-q_0-3q_0^2+j_0 \right] z^2
\nonumber
\\
\lo{+}
{1\over24}\left[
2-2q_0-15q_0^2-15q_0^3+5j_0+10q_0j_0+s_0 
\right] z^3
+ O(z^4) \Bigg\}.
\end{eqnarray}
It is, however, important to realise that this is \emph{not the same}
as saying that cosmological inflation predicts $k=0$. Instead what
generic cosmological inflation predicts is the much weaker statement
that for all practical purposes the \emph{present day} universe is
indistinguishable from a $k=0$ spatially flat universe.  This means
that if our universe happens to be a topologically trivial $k=0$ FRW
cosmology, then in a formal logical sense \emph{we will never be able
  to prove it}.  All we will ever be able to do is to place
increasingly stringent lower bounds on the dimensionless parameter
$H_0\, a_0/c$, but this will never rigorously permit us to conclude
that $k=0$. The fundamental reason for this often overlooked but
trivial observation is that a topologically trivial $k=0$ FRW universe
can be mimicked to arbitrary accuracy by a $k=\pm1$ FRW universe
provided the scale factor is big enough. (If the universe has
nontrivial spatial topology there is a possibility of using the
compactification scale, which might be [but does not have to be] much
smaller than the scale factor, to indirectly distinguish between
$k=-1/0/+1$.) In contrast if the true state of affairs is $k=\pm1$,
then with good enough data on $H_0 \,a_0$ we will in principle be able
to determine upper bounds which (at some appropriate level of
statistical uncertainty) demonstrate that $k\neq0$.  Also note that
even in inflationary cosmologies it is not true that $H(t) a(t)/c \gg
1$ at \emph{all} times, and in particular this inequality \emph{may}
be violated (and very often is violated) in the pre-inflationary
epoch.

\section{Discussion}

The presentation of this article now makes it clear what can and
cannot be expected, even in principle, from improved observations of
the luminosity distance Hubble law $d_L(z)$ (or for that matter its
angular-distance or proper-motion distance variants).  As more data
is collected, at progressively higher redshifts, we can better bound
the derivatives $\d^n [d_L(z)]/\d z^n|_{z=0}$. This allows one in
principle to extract the Hubble and deceleration parameters, but even
at $O(z^3)$ there is a problem in that the number of free parameters
in the Hubble law exceeds the number of measurable coefficients. This
arises because of the fact that the $O(z^3)$ term depends explicitly
on both the jerk $j_0$ and the scale factor $a_0$. This problem
persists at $O(z^4)$ and higher, with the number of free parameters in
the Hubble law always exceeding the number of measurable coefficients
by one.

If (\emph{and only if}) one has some independent method for bounding the
scale factor $a_0$ (or more precisely the space curvature $k/a_0^2$)
can one even in principle use the observational Hubble law to bound
the jerk, snap, and higher time derivatives of the scale factor.  One
could for instance use theoretical considerations based on the assumed
occurrence of cosmological inflation to effectively set $k/a_0^2\to0$,
but should then be aware that one is making a very definite additional
theoretical assumption~\cite{jerk-and-snap,Riess,Caldwell}, well
beyond the simple symmetry considerations of the standard FRW
cosmology.

It is also worth noting that nothing in this article has made any use
of the Einstein equations, or their specialization to FRW spacetimes,
the Friedmann equations. Thus all comments made in this article are
completely independent of one's favourite choice of matter model for
the cosmological fluid.  There are currently very many quite radically
different models for the cosmological fluid under active
consideration. Though these models often make dramatically differing
predictions in the distant past (\eg, a ``bounce'') or future (\eg, a
``big rip'') there is considerable degeneracy among the models in that
many physically quite different models are compatible with present day
observations.  Despite the fact that some parameters in cosmology are
now known to high accuracy, other parameters can still only be crudely
bounded~\cite{precision}. In particular, published bounds on the jerk
parameter are still relatively weak, and published bounds on the snap
parameter are nonexistent~\cite{jerk-and-snap,Riess,Caldwell}.

\appendix

\ack

This Research was supported by the Marsden Fund administered by the
Royal Society of New Zealand.

\section*{References}


\end{document}